\documentclass[aps,twocolumn,showpacs,preprintnumbers,amsmath,amssymb]{revtex4}
\usepackage{graphicx}
\usepackage{dcolumn}
\usepackage{bm}

\newcommand{\bef}{\begin{figure}}
\newcommand{\eef}{\end{figure}}
\newcommand{\nn}{\nonumber}
\newcommand{\be}{\begin{equation}}
\newcommand{\ee}{\end{equation}}
\newcommand{\bea}{\begin{eqnarray*}}
\newcommand{\eea}{\end{eqnarray*}}
\newcommand{\ra}{\rightarrow}
\newcommand{\gm}{\gamma^\mu}
\newcommand{\gf}{\gamma^5}
\newcommand{\N}{\bar{N}}
\newcommand{\del}{\partial}
\newcommand{\bt}{{\boldsymbol{\tau}}}
\newcommand{\ba}{{\boldsymbol{a_1}}}
\newcommand{\br}{{\boldsymbol{\rho}}}
\newcommand{\bp}{{\boldsymbol{\pi}}}
\newcommand{\cl}{{\cal{L}}}
\begin{document}

\title{Thermal Radiation from Nucleons and Mesons}

\author{Jan-e Alam$^1$, Pradip Roy$^2$ and Sourav Sarkar$^1$}

\medskip

\affiliation{$^1$Variable Energy Cyclotron Centre, Kolkata 700064, India\\
$^2$Saha Institute of Nuclear Physics, Kolkata 700064, India}

\date{\today}

\begin{abstract}
Thermal photon emission rates due to meson-nucleon interactions have
been evaluated. An exhaustive set of reactions involving $p(\bar p)$, 
$n(\bar n)$, $\rho$, $\omega$, $a_1$, $\pi$ and $\eta$ is seen to
provide a sizeable contribution to the emission rate from hot hadronic 
matter.  Contributions from baryonic resonances are found to be negligibly 
small.
\end{abstract}

\maketitle

With results coming in from Au-Au collisions at the Relativistic Heavy
Ion Collider (RHIC) at Brookhaven National Laboratory (BNL), the
search for signals of quark gluon plasma~(QGP) is being pursued with
renewed vigour. It is well known that electromagnetic signals are the only
ones which probe the entire space-time history of the collisions as
they are emitted at all stages and undergo very little re-scattering.
Since signals from QGP have to be unearthed from the huge debris of
radiations from hadronic sources it becomes all the more necessary
to have an accurate estimation of the same. Apart from this the emission
of photons from the hadronic matter is also important to investigate
the modifications of the hadronic spectral functions in a thermal bath
~\cite{annals}.

The hadronic matter is usually considered to be a gas of the
low lying mesons $\pi$, $\rho$, $\omega$ and $\eta$. Reactions between these
as well as the decays of the $\rho$ amd $\omega$ were considered~\cite{old1}
to be the source of thermal photons from hadronic matter.
In~\cite{old2} the role of the intermediary $a_1$ meson 
in the $\pi\rho\ra\pi\gamma$ channel was discussed. 
We~\cite{our_npa} found an enhancement in the photon production by
incorporating medium effects through thermal loop corrections on the 
hadronic decay widths and masses. In our quest for additional
sources of thermal photon production we failed to find discussions on
the role of baryons and {\em a priori} we find no reason for this.
We hence consider it worthwhile to investigate, in particular,
the role of meson-nucleon interactions for photon emission. 
In an earlier work~\cite{prc03} we had found
the contribution due to the decay of baryon 
resonances to be small.

In this work we evaluate photon emission due to nucleon (and antinucleon)
scattering from $\pi$, $\rho$, $\omega$, $\eta$ and $a_1$ mesons
in the thermal bath.
Listed below are the standard phenomenological interactions 
which have been used:
\begin{widetext}
\bea
\cl_{VNN}&=&g_{\rho NN}\,\left[\N\gm\bt N\cdot\br_\mu-\frac{\kappa}{2m_N}\,
\N\sigma^{\mu\nu}\bt N\cdot\del_\nu\br_\mu \right]
-g_{\omega NN}\,\N\gm N\omega_\mu\nn\\
\cl_{ANN}&=&\frac{g_{\pi NN}}{m_\pi}m_{a_1}\,\N\gf\gm\bt N\cdot\ba_{\mu}\nn\\
\cl_{PNN}&=&\frac{g_{\pi NN}}{m_\pi}\,\N\gf\gm\bt N\cdot\del_{\mu}\bp
+\frac{g_{\eta NN}}{m_\eta}\,\N\gf\gm N\del_{\mu}\eta\nn\\
\cl_{em}&=&eA^\mu\left[\N\gamma_\mu N-[\bp\times\del_\mu\bp
-\br^\nu\times(\del_\nu\br_\mu-\del_\mu\br_\nu)]_3\right]
+\frac{e}{2}F^{\mu\nu}(\br_\mu\times\br_\nu)_3
+g_{a_1\pi\gamma}\ba_\mu\cdot\del_{\nu}\bp\,F^{\mu\nu}\nn\\
\eea
\end{widetext}
Here, $N$ represents the nucleon isospin doublet 
$\left(\begin{array}{c}p\\n\end{array}\right)$, 
$F_{\mu\nu}=\del_\mu A_\nu-\del_\nu A_\mu$ and $\bt$ are 
the Pauli matrices. We have included monopole form factors at
the strong vertices. They have the generic form 
$(\Lambda^2-M^2)/(\Lambda^2-X^2)$
where $\Lambda$ and $M$ stand for the cutoff and the mass of the exchanged
particle in the $X(=t/u)$ channels respectively. The values of the
cutoff and the coupling constants have been  taken from
Ref.~\cite{roper} and are listed in Table 1. 
It is to be noted that the number of Feynman diagrams involved
in the present calculations with all the isospin combinations 
of mesons and baryons is very large. The invariant amplitudes
for these diagrams will require too large a space to be presented here.

In Fig.~\ref{fig1}, we plot the rate of photon emission from a thermal hadronic 
medium consisting of mesons and baryons at $T$=170 MeV and zero
baryonic chemical potential.
For convenience we have subdivided the large number of photon
producing reactions into three categories involving nucleons
and pseudoscalar mesons ($P$=$\pi$, $\eta$), vector mesons 
($V$=$\rho$, $\omega$ and $\phi$) and axial vector ($A$=$a_1$) meson. Type
$P$ consists of reactions of the kind $P N\ra\gamma N$ and $N\N\ra\gamma P$,
$V$ of the type $V N\ra\gamma N$ and $N\N\ra\gamma V$ and 
$A$ of the type $A N\ra\gamma N$ and $N\N\ra\gamma A$. Due to the the
relatively high abundance of pions in the system, type $P$ has the 
highest contribution at lower energies. $V$ type reactions are found
to dominate after about 1.5 GeV. The important factor in this case is
the large spin-isospin degeneracy of the $\rho$ meson. Also shown is
the contribution to the photon emission rate from the decay of 
heavy baryons $R$ to nucleons, $N$ and photons, $R\ra \gamma N$. 
Here $R$ consists of the baryonic
resonances N(1440), N(1520), N(1535), N(1650), N(1675), N(1680),
N(1700), $\Delta(1232), \Delta(1620), \Delta(1700), \Delta(1905)$ and
$\Delta(1950)$. 
The partial decay widths of $R$ have been taken from Ref.  ~\cite{PDG}.

\renewcommand{\arraystretch}{1.5}
\vskip 0.2in
\begin{center}
\begin{table}
\begin{tabular}{cccc}
\hline
\hline
Vertex & Process & Coupling & $\Lambda$ (GeV) \\
\hline
$NN\pi$ & $\pi$ exchange & $\frac{g_{\pi NN}^2}{4\pi}$=0.079 & 0.6 \\
 & $N$ exchange &  & 1.3 \\
$NN\eta$ & $N$ exchange & $\frac{g_{\eta NN}^2}{4\pi}$=0.148 & 2.5 \\
$NN\rho$ & $\rho$ exchange & $\frac{g_{\rho NN}^2}{4\pi}$=0.84 & 1.4 \\
 &  & $\kappa$=6.1 &  \\
 & $N$ exchange &  & 1.2 \\
$NN\omega$ & $N$ exchange & $\frac{g_{\omega NN}^2}{4\pi}$=20.0 & 1.2 \\
$a_1\pi\gamma$ &  & $g_{a_1\pi\gamma}$=0.743 & \\
\hline
\end{tabular}
\caption{Values of the coupling constants and cutoff parameters used.}
\end{table}
\end{center}

In fig.~\ref{fig2} the emission rates from the 
baryonic degrees of freedom have been shown at
a temperature 140 MeV. The introduction of the monopole
form factors at the meson nucleon vertices reduces the 
the production rate by a factor of 3 around $E_\gamma\sim 2$
GeV.  A similar kind of suppression (see first reference of ~\cite{old1}) 
has been observed with the inclusion of form factors at the interaction 
vertices in the emissivity  from the  reactions involving only mesons
($\pi$,\,$\rho$,\,$\omega$\, $\eta$ and $a_1$). A factor of $\sim 3$
suppression of the production rate of photons from a mesonic gas
will make it comparable with that 
from the $P$, $V$ and $A$ types of  reactions defined before.

\bef
\begin{center}
\includegraphics[scale=0.45]{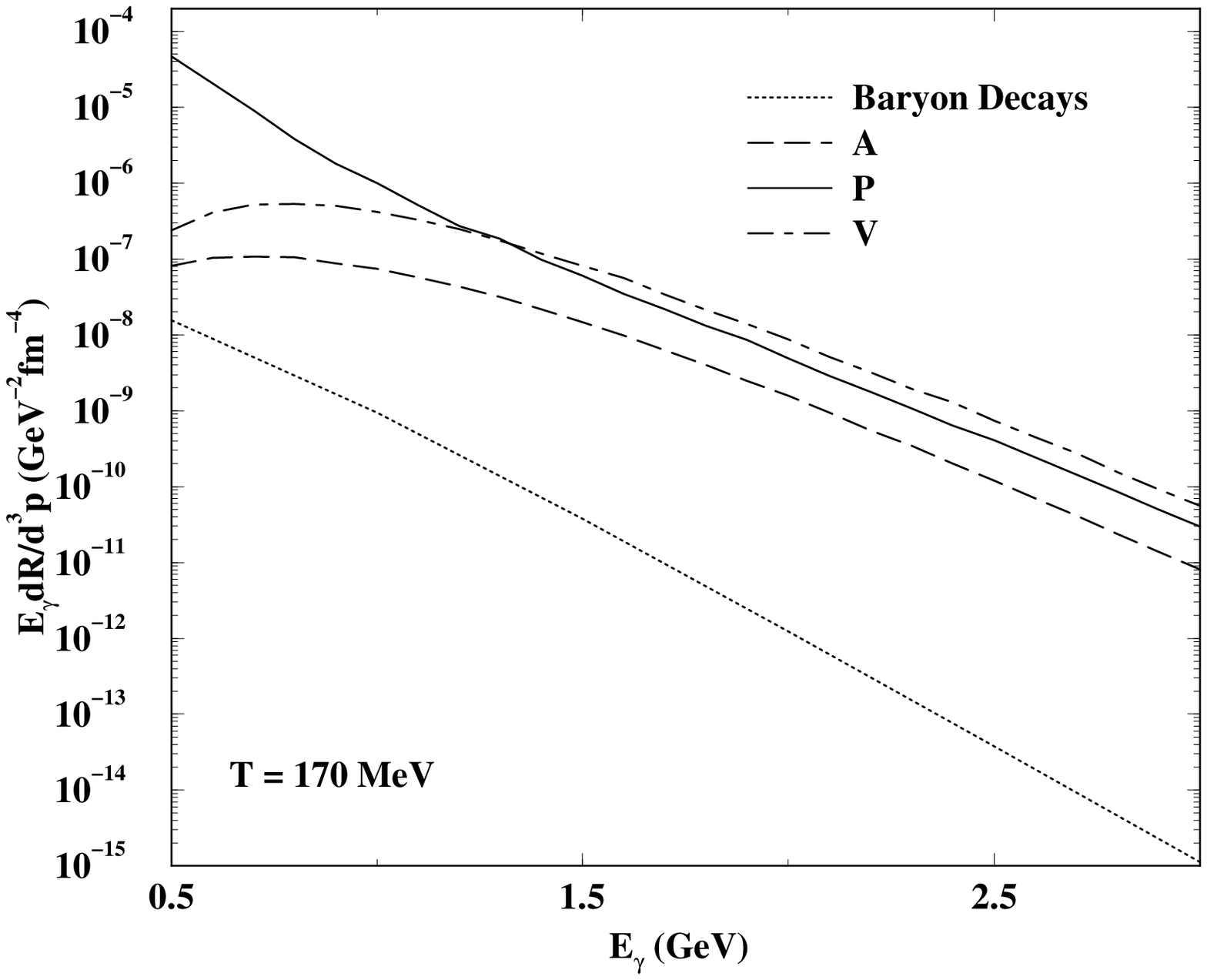}
\caption{Thermal photon emission rates from meson-nucleon interactions at
$T$=170 MeV. The legends $P$, $V$ and $A$ are described in the text. 
}
\label{fig1}
\end{center}
\eef
\bef
\begin{center}
\includegraphics[scale=0.45]{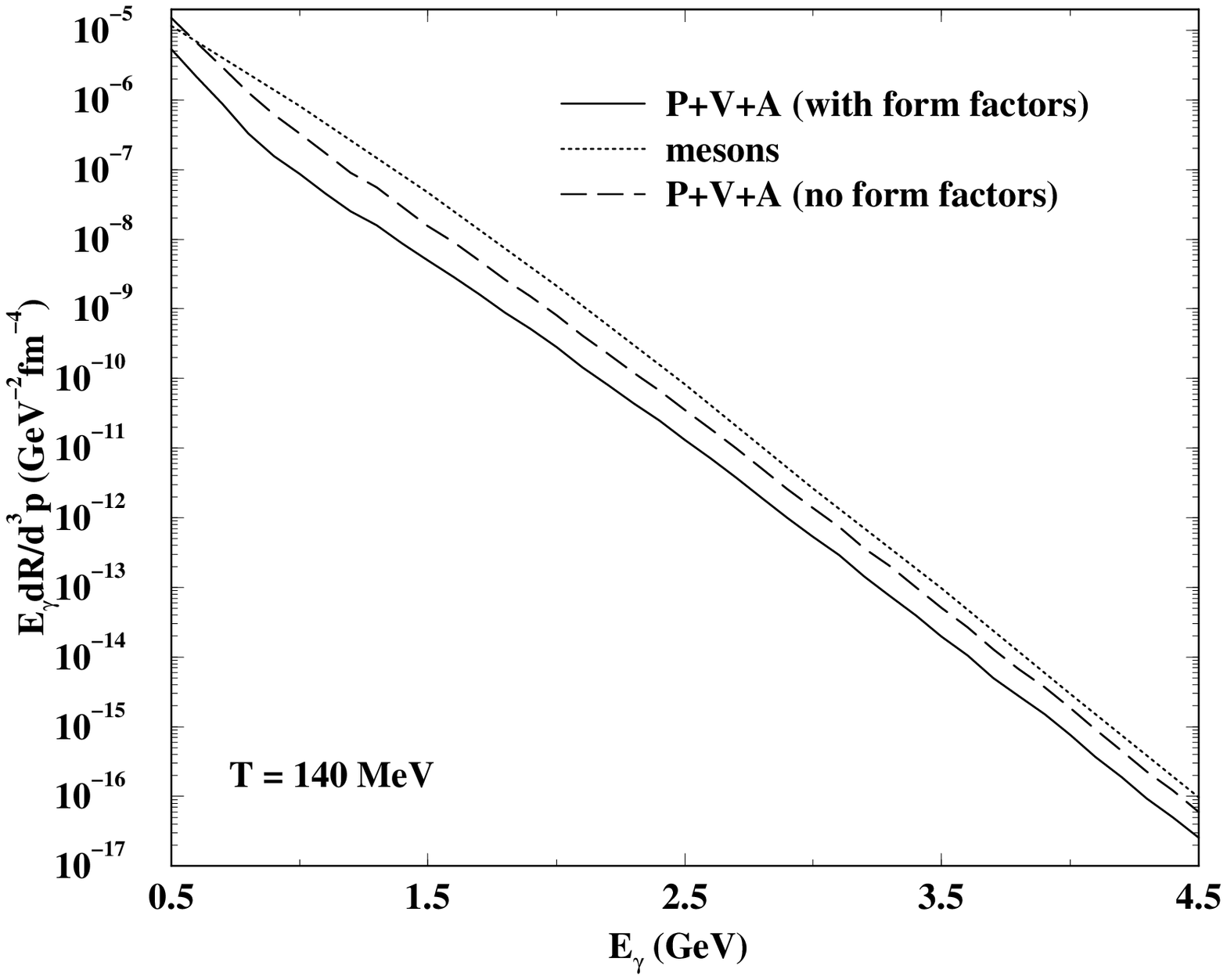}
\caption{Thermal photon emission rates at $T$=140 MeV. 
Solid and dashed line indicate emission rate from a hadronic
gas involving nucleons with and without form factors respectively. 
The parameters of the form factors are given in Table I. 
The dotted line represents result from a mesonic gas of 
$\pi$,\,$\rho$,\,$\omega$\, $\eta$ and $a_1$.
}
\label{fig2}
\end{center}
\eef
\bef
\begin{center}
\includegraphics[scale=0.45]{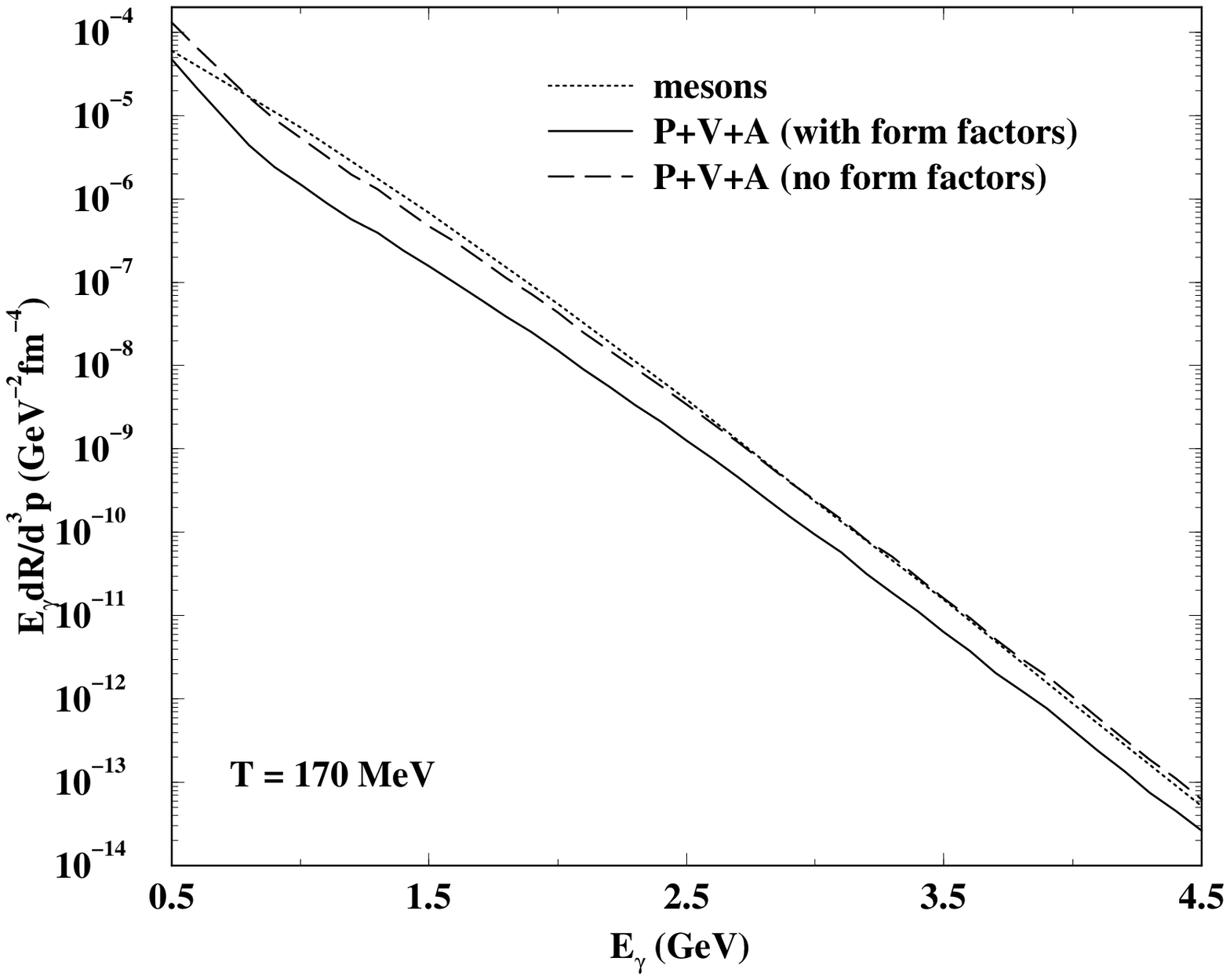}
\caption{Same as Fig.~\protect{\ref{fig2}} at $T$=170 MeV. 
}
\label{fig3}
\end{center}
\eef

The emission rates of photons from $P$, $V$ and $A$ types of 
reactions will become important with increase in temperature. In fact,
the production rates with only mesonic degrees of freedom is comparable
to the corresponding rates from $P$, $V$ and $A$ type of reactions if
the effects of the vertex form factors are ignored (fig.~\ref{fig3}). 
Considering the fact that the form factors at the 
interaction vertices suppress the emission rates from 
mesons as well as from reactions involving nucleons, 
the production rates involving nucleons are not negligible. 
 
In summary, we have evaluated the production rate of photons
from an exhaustive set of reactions involving mesons and
baryons. In all the earlier works, the role of the baryons
on the photon emission rates were neglected, though it
appears to be important.  We observe 
that the production rates from the reactions involving nucleons
are suppressed primarily because of the inclusion of the form factors
at the meson-nucleon vertices. At non-zero baryonic chemical
potential the emission rates from the reactions involving
nucleonic degrees of freedom may be enhanced. However, for
very high colliding energies such as at RHIC the value
of the chemical potential is expected to be small.

\end{document}